# Generalizing Deep Whole Brain Segmentation for Pediatric and Post-Contrast MRI with Augmented Transfer Learning

Camilo Bermudez*[a], Justin Blaber[b], Samuel W. Remedios[c], Jess E. Reynolds[d], Catherine Lebel[d], Maureen McHugo[e], Stephan Heckers[e], Yuankai Huo[b], Bennett A. Landman[a,b]

[a] Department of Biomedical Engineering, Vanderbilt University, 2201 West End Ave, Nashville, TN, USA 37235; [b] Department of Electrical Engineering, Vanderbilt University, 2201 West End Ave, Nashville, TN, USA 37235; [c] Center for Neuroscience and Regenerative Medicine, Henry Jackson Foundation, 6720A Rockledge Dr, Bethesda MD 20817; [d] Department of Radiology, University of Calgary, 28 Oki Dr, Calgary, Alberta, Canada; [e] Department of Psychiatry, Vanderbilt University Medical Center; 1211 Medical Center Dr, Nashville, TN, USA 37235

**ABSTRACT**

Generalizability is an important problem in deep neural networks, especially in the context of the variability of data acquisition in clinical magnetic resonance imaging (MRI). Recently, the Spatially Localized Atlas Network Tiles (SLANT) approach has been shown to effectively segment whole brain non-contrast T1w MRI with 132 volumetric labels. Enhancing generalizability of SLANT would enable broader application of volumetric assessment in multi-site studies. Transfer learning (TL) is commonly used to update the neural network weights for local factors; yet, it is commonly recognized to risk degradation of performance on the original validation/test cohorts. Here, we explore TL by data augmentation to address these concerns in the context of adapting SLANT to anatomical variation and scanning protocol. We consider two datasets: First, we optimize for age with 30 T1w MRI of young children with manually corrected volumetric labels, and accuracy of automated segmentation defined relative to the manually provided truth. Second, we optimize for acquisition with 36 paired datasets of pre- and post-contrast clinically acquired T1w MRI, and accuracy of the post-contrast segmentations assessed relative to the pre-contrast automated assessment. For both studies, we augment the original TL step of SLANT with either only the new data or with both original and new data. Over baseline SLANT, both approaches yielded significantly improved performance (signed rank tests; pediatric: 0.89 vs. 0.82 DSC, p<0.001; contrast: 0.80 vs 0.76, p<0.001). The performance on the original test set decreased with the new-data only transfer learning approach, so data augmentation was superior to strict transfer learning.

**Keywords:** Transfer learning, whole brain segmentation, generalizability, magnetic resonance imaging

## 1. INTRODUCTION

Whole brain segmentation provides a quantitative, noninvasive measurement of neuroanatomy from magnetic resonance imaging (MRI). Although a manual delineation of brain structures is considered to be the gold standard, this process can be time- and resource-intensive [1]. For many years, atlas-based segmentation was considered the standard for automatic whole brain segmentation due to high accuracy across over 100 labels [1-3]. However, a key limitation of atlas-based segmentation is the computational time required, which limited the ability to analyze large-scale cohorts [1, 4]. Until recently, a full-resolution, whole brain segmentation with over 100 labels was a challenging task due to hardware constraints. However, this problem was solved by Huo et al. by using a tile-based method called the spatially localized atlas network tiles (SLANT) to divide T1-weighted brain MRI into 27 overlapping tiles and train independent networks, which are then fused via majority voting [5, 6]. This method showed better performance than multi-atlas and other deep learning-based methods on brain MRI acquired for research. Unfortunately, when evaluated on same-subject clinical data, investigators found a high coefficient of variation between volume estimates of several ROIs across clinical T1 MRI acquisition modalities within the same subject for pre- and post-contrast imaging [7]. This observation highlights a key obstacle in deep learning-based image processing tools: most algorithms are trained and tested on high-quality research scans (e.g., high resolution and SNR) and reported accuracy reflects performance on research data. However, rarely is there a description of performance on clinical data, which is largely heterogenous due to resolution, noise, artifact, or the presence of intravenous contrast. For instance, Figure 1 shows T1w brain MRI from three different scenarios: one adult subject acquired for research, one pediatric subject acquired for research, and one clinically acquired adult subject with intravenous contrast. These images highlight the heterogeneity in clinical data due to differences in anatomy, contrast, or noise. It is unclear how deep learning algorithms trained on research data will generalize to heterogeneous clinical datasets.

One common approach towards generalizability across sites and scanners is to use transfer learning (TL) to update the neural network weights for local factors [8, 9]. For instance, if SLANT were applied at a new site, its performance could be refined by taking a number of training examples and updating the weights of the original network. However, this technique is commonly recognized to risk degradation of performance on the original validation/test cohorts, thus limiting its generalizability [8]. Nevertheless, it is desirable to improve performance on datasets that deviate from the original training data. For instance, it has been shown that image processing algorithms have decreased performance in young children due to differences in gray matter to white matter contrast and limited training examples [10]. Similarly, the presence of intravenous contrast may affect the performance of intensity-based methods such as deep neural networks [7]. Although most algorithms are trained on adult images exclusively with or without intravenous contrast, it would be ideal to generate reproducible results in imaging despite the presence of contrast. Pediatric and contrast-enhanced data are to examples of common, but underrepresented data in deep learning algorithms.

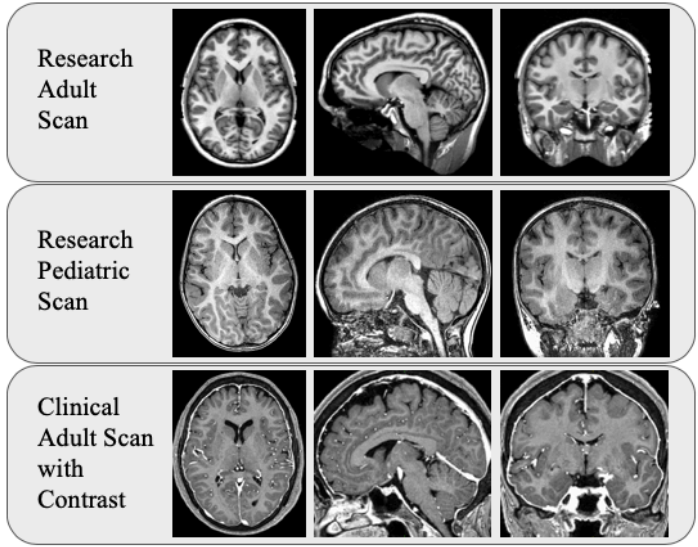

**Figure 1.** Many modern algorithms in medical image processing are developed using magnetic resonance (MR) scans with high resolution and SNR. It is unclear how these algorithms translate to noisier, more heterogeneous data acquired during clinical practice. Here, we compare a research scan from the OASIS dataset to two clinically acquired brain MR imaging: a pediatric subject and an adult subject. These images highlight the heterogeneity in clinical data due to differences in anatomy, contrast, or noise. In this work, we propose a solution to generalize an existing segmentation algorithm to these circumstances.

In this work, we explore how properties in imaging datasets may affect performance during TL. We hypothesize that although deep neural network performance can be refined with TL in order to fit specific dataset properties, this may result in reduced performance on the original training data. The novelty of this work is a proposed solution of augmented TL to preserve generalizability and performance in the whole brain segmentation task. We evaluate this hypothesis on a cohort of pediatric research participants and a cohort of clinically acquired data with intravenous contrast. We compare our performance with an adult imaging dataset acquired for research purposes.

## 2. METHODS

### 2.1 Imaging Datasets

The original SLANT whole brain segmentation algorithm leverages a large dataset of automatically generated labels (N = 5,111 subjects) to pretrain the model before refining through TL using a small dataset of manually labelled ground truth atlases [5, 6]. In this work, we begin with the pretrained model and use three different imaging datasets to test the effect of TL on algorithm generalizability. We assess the performance of TL with three different datasets: 1) An adult T1-weighted brain MRI dataset with manual labels, 2) A pediatric T1-weighted brain MRI dataset with manually corrected labels, and 3) A paired clinical dataset with pre- and post-contrast brain MRI without manual labels.

The original imaging dataset used for TL in SLANT consists of 45 subjects with T1-weighted brain MRI from the Open Access Series on Imaging Studies (OASIS) [11], with ages 18-96 years old. These scans were acquired at 3T field strength at a field of view which varies from 256x270x256 to 256x334x256 all with 1mm isotropic resolution. Each scan has 133 manual labels traced according to the BrainCOLOR protocol used as ground truth [12]. Accuracy was measured using the Dice similarity coefficient (DSC) between the predicted whole brain segmentation and manual labels [13].

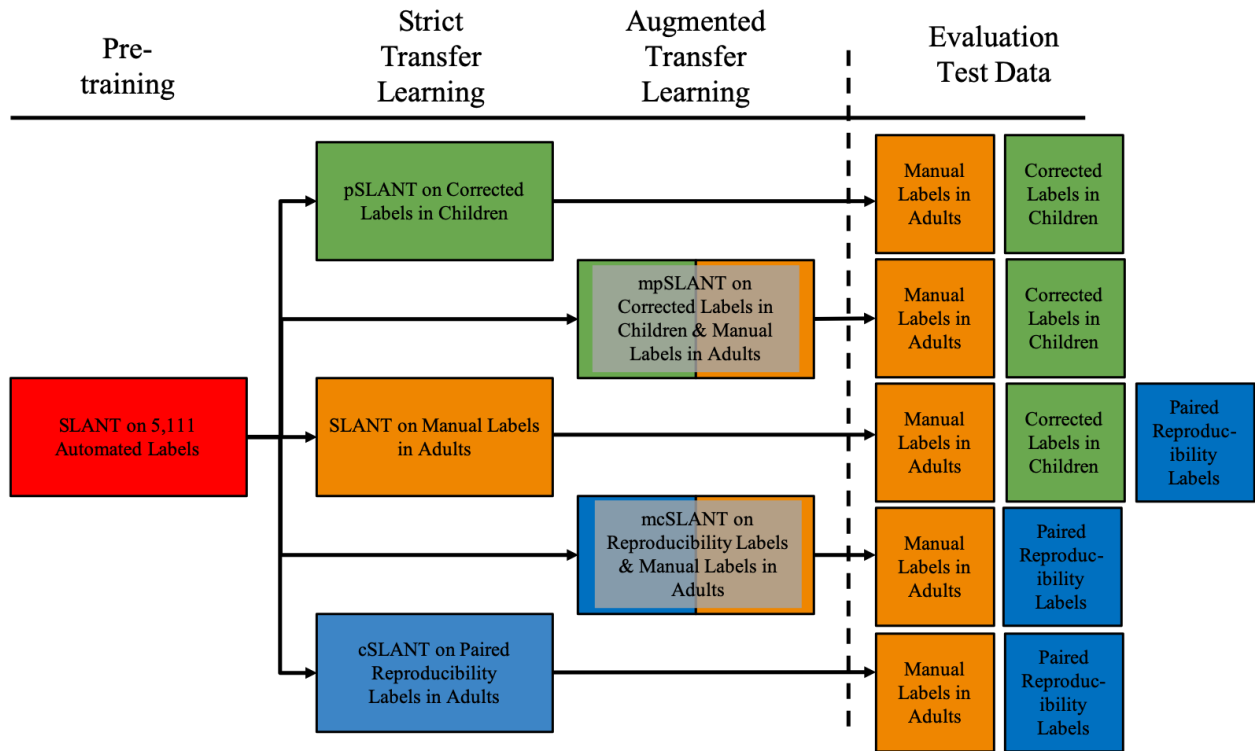

**Figure 2.** We explore transfer learning by data augmentation to address concerns of generalizability in the context of adapting SLANT whole brain segmentation to anatomical variation (e.g., adults versus children) and scanning protocol (e.g., non-contrast research T1w MRI versus contrast enhanced clinical T1w MRI). We begin with a model that has been pretrained on 5,111 subjects with a whole brain segmentation obtained from multi-atlas segmentation. Next, we consider three datasets for transfer learning: First, we identified 30 T1w MRI of children with manually corrected volumetric labels, and accuracy of automated segmentation was defined relative to the manually provided truth. Second, we used the original 45 manually traced atlas from adult brains used in SLANT to refine the initial pretrained segmentation. Third, we retrieved 36 paired datasets of pre- and post-contrast clinically acquired T1w MRI, and accuracy of the post-contrast segmentations were assessed relative to the pre-contrast automated assessment.

The second dataset consists of 30 pediatric subjects (ages 2.34 – 4.31 years old) with T1-weighted brain MRI. Imaging was acquired at 3T field strength with field of view 256x210x256 at 0.9 mm isotropic resolution. Initial automatic whole brain segmentation was obtained into the same 133 regions using multi-atlas segmentation from 35 adult atlases as outlined by [14-16]. These labels were then manually corrected by a rater expert in pediatric neuroimaging. Accuracy of segmentation is measured using DSC between the manually corrected labels and the output of the automatic segmentation.

The third dataset consists of 36 subjects with clinically acquired, paired T1-weighted brain MRI with and without intravenous contrast. These subjects were acquired with 3T field strength and a field of view of 256x170x256 at 1mm isotropic resolution. Paired images were co-registered using NiftyReg affine registration [17]. Labels for the same 133 regions were generated for the pre-contrast image and used as labels for the post-contrast image using the original SLANT algorithm.

### 2.2 Transfer Learning Pipeline

Throughout this work, we begin all of our experiments with a SLANT model of whole brain segmentation that has been pretrained in 5,111 subjects using automatically generated manual labels, as described by Huo et al [5]. In this work, Huo performs TL using the dataset of 45 adults with manual labels for best results [5]. Here, we use this result from Huo et. al as the baseline comparison. We seek to train a network that achieves high performance in a new, unseen dataset while preserving performance on the original adult cohort. Figure 2 outlines the proposed pipeline from pre-training to TL and test set evaluation.

In the first experiment, we wish to refine the segmentation for a pediatric population. We perform TL to train a model using the pediatric cohort exclusively called pediatric SLANT (pSLANT) as well as an augmented TL using a mixed training cohort of adults and pediatric subjects called mixed pediatric SLANT (mpSLANT). Here, we aim to maximize the performance DSC (pDSC), or the accuracy relative to the manually edited labels, while preserving the performance on the adult cohort.

In the second experiment, we aim to refine the whole brain segmentation to be consistent despite the presence of intravenous contrast. We use TL to train a model using the contrast-enhanced images exclusively called cSLANT as well as an augmented TL using a mixed training cohort that includes contrast enhanced images as well as the adult subjects with manually traced labels (mcSLANT). Here, we seek to maximize reproducibility DSC (rDSC), or the accuracy between the pre-contrast and post-contrast image, which represents similarity of automatic segmentation between two images of the same subject.

### 2.3 Performance Analysis

All models were trained with 80% of each of the subjects as the training set, while the remaining 20% was evenly split between validation and testing. Model accuracy was evaluated on the validation set using DSC, with the final model chosen as having the highest validation set DSC after 30 epochs of TL as in [5, 6]. A five-fold cross-validation scheme was employed in independently trained models. All models were compared using Wilcoxon signed-rank test with Bonferroni corrections on a per experiment/figure basis (n=3 to 6). All experiments were performed on an NVIDIA GForce Titan GPU with 12 GB memory. The original SLANT algorithm as well as TL modifications were implemented in PyTorch v0.4 [18]. TL training was performed with the Adam optimizer using the Dice coefficient loss function at a learning rate of 0.0001 [13, 19].

## 3. RESULTS

### 3.1 Transfer Learning on Pediatric Imaging

The first set of experiments seeks to improve performance on the whole brain segmentation of a pediatric population. As a baseline, we perform a prospective evaluation of the original SLANT algorithm on this dataset, which achieves a mean DSC of 0.82 +/- 0.01. Figure 3 shows the best performance on the pediatric dataset is achieved when the TL dataset is replaced entirely by the cohort of 30 pediatric subjects (pDSC = 0.90 +/- 0.01, p < 0.001), followed closely by an augmented TL dataset comprised of both adult and pediatric subjects (pDSC = 0.89 +/- 0.01). Interestingly, pDSC performance on the withheld sample from the adult dataset dropped in both cases from 0.715 +/- 0.054 to 0.698 +/- 0.061 (p < 0.001) when training on the pediatric dataset exclusively and to 0.704 +/- 0.065 (p < 0.001) with a mixed training sample (Figure 5).

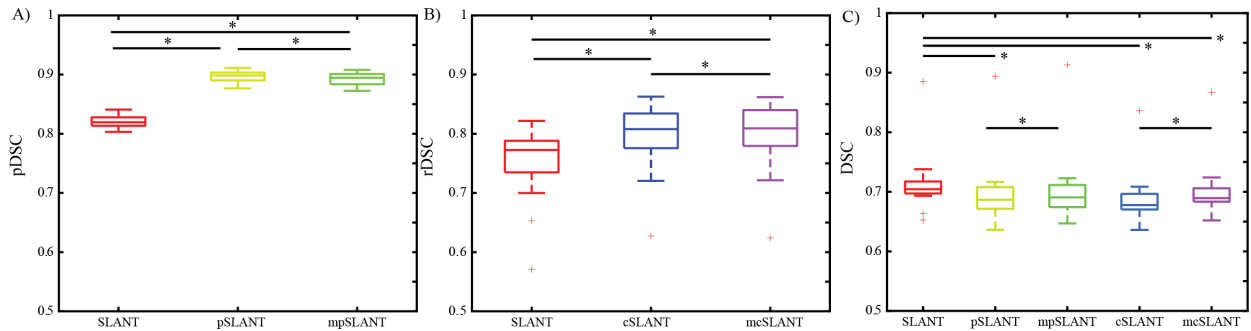

**Figure 3. Quantitative results of augmented transfer learning.** A) The best performance DSC (pDSC) on the pediatric dataset was achieved when using the pediatric cohort exclusively during transfer learning showing an increase of 0.08 in pDSC compared to the original SLANT. Statistical significance (*) is achieved at p < 0.0167 (Wilcoxon Sign-Rank Test with Bonferroni corrections). B) The best reproducibility DSC (rDSC) is achieved when using a mixed cohort of contrast-enhanced images and manually traced non-contrast images, for a mean increase in rDSC of 0.05 compared to original SLANT. Statistical significance (*) is achieved at p < 0.0167 (Wilcoxon Sign-Rank Test with Bonferroni corrections). C) There is a decrease in performance of the manual labels in the adult dataset when new data are introduced, although this decrease is mitigated when using a mixed training cohort for both pediatric SLANT and contrast SLANT. Statistical significance is achieved at p < 0.0083 (Wilcoxon Sign-Rank Test after Bonferroni corrections).

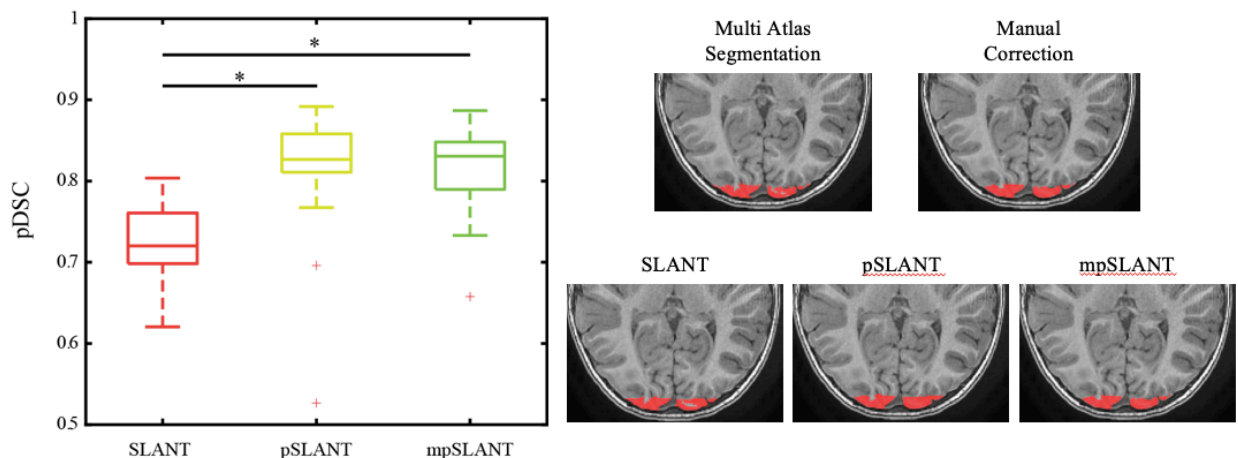

**Figure 4. Qualitative performance of pediatric SLANT on the occipital pole labels.** Here we show the performance of transfer learning on the pediatric dataset on the occipital pole. This region was chosen since it was the region that required the most extensive manual editing. Performance is compared to the manually corrected labels. Here, we see that both pSLANT and mpSLANT show a dramatic improvement in performance compared to the original SLANT. For comparison, we have the DSC between the original automatic multi-atlas segmentation labels and the manual corrections. On the right, we see the segmentation overlaid on brain MRI for all methods. Statistical significance (*) is achieved at p < 0.0167 (Wilcoxon Sign-Rank Test with Bonferroni corrections).

An important goal of this task is to enforce the modifications introduced during manual correction of the automatic labels from multi-atlas segmentation in order to introduce expert context. Huo et al. showed that the mean performance for multi-atlas segmentation is DSC of 0.760 +/- 0.012 on the adult cohort. When this method is applied to the pediatric cohort, the DSC between the generated labels from multi-atlas segmentation and the subsequent edits was 0.912 +/- 0.006. We examine the bilateral occipital poles as a region of interest (ROI) that underwent extensive manual editing as a representative ROI for this task. Figure 4 shows that the mpSLANT model trained on pediatric data or a mixed adult and pediatric dataset results in higher DSC relative to the original SLANT. Figure 4 also shows labels and output segmentation overlaid on a representative subject with median performance.

### 3.2 Transfer Learning on Clinical Imaging with Intravenous Contrast

The second set of experiments seek to use paired imaging to enforce consistency in segmentation within subjects despite the presence of intravenous contrast. We use labels generated on the pre-contrast image as truth to train a segmentation on the contrast-enhanced image. Figure 4 shows that a baseline prospective performance of the original SLANT on the occipital loves achieves a mean rDSC of 0.757 +/- 0.048 between pre- and post-contrast images. The best performance is achieved when using an augmented TL dataset of both manually traced labels and pre-contrast labels (rDSC = 0.803 +/- 0.048, p<0.001) followed by TL with pre-contrast labels exclusively (rDSC = 0.799 +/- 0.047). When evaluating performance on the original adult dataset, there is a drop in DSC from 0.715 +/- 0.054 to 0.688 +/- 0.047 (p< 0.001) when training with the paired dataset exclusively. When training with a mixed dataset, the performance on the adult dataset shows a reduced drop in DSC to 0.701 +/- 0.050 (p < 0.001).

To further validate the reproducibility of our method, we examine the differences in measured volume of the hippocampus, a small, gray matter structure in the temporal lobe, in the pre- and post-contrast images. Due to its involvement in learning and memory [20], the hippocampus structure has been widely studied as a potential imaging biomarker for Alzheimer's Disease [21], psychosis[22], and epilepsy [23], thus making it a desirable target for reproducible volumetric measurements across scans within the same subject [14, 24, 25]. Figure 5 shows that using the original SLANT, there is an average 9.91% decrease in volume of the hippocampus in the contrast-enhanced image within the same subject (RMSE = 0.512 cm$^3$). However, the cSLANT model trained with a mixed dataset of manually traced adults and pre-contrast labels, shows only a 0.89% average decrease in volume from the pre-contrast to the post-contrast image (RMSE = 0.250 mm$^3$). The output segmentation for each method is overlaid on the pre- and post-contrast images in Figure 5.

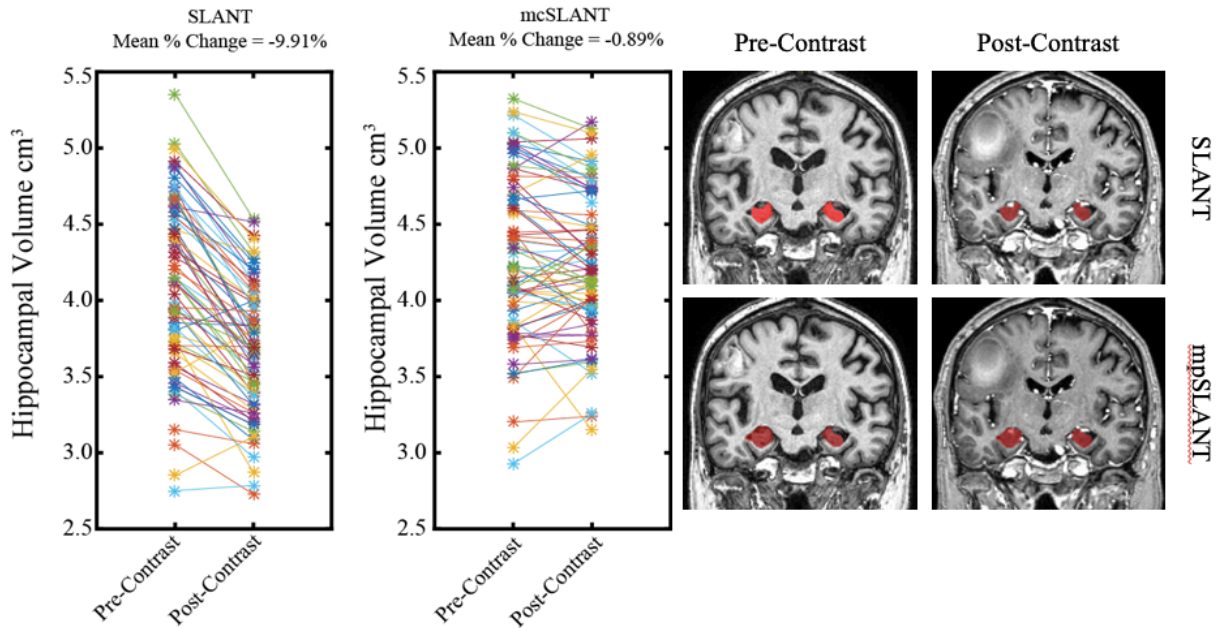

**Figure 5. Qualitative performance of contrast SLANT on the hippocampus volume.** We see that using original SLANT, there is a decrease in estimated volume of about 10%, compared to an estimated change in hippocampus volume of less than 1% in the case mcSLANT. The label overlays on the right show the output segmentation for each method overlaid on the brain MRI.

## 4. DISCUSSION & CONCLUSIONS

In this work, we explore the effect of TL on the performance of deep neural networks. We were able to refine the existing whole brain segmentation algorithm SLANT with new datasets. We showed that the original SLANT segmentation algorithm has decreased performance in pediatric brains, likely due to smaller volume and altered gray/white matter proportions in the younger subjects compared to the initial training data used in SLANT. However, we can improve performance by 0.07 DSC when we introduce pediatric subjects in the TL process in addition to the original adult subjects. We showed that segmentation of difficult regions, such as the occipital poles, can be improved by providing mixed training examples. Importantly, we showed that the proposed method can reproduce the manual corrections done on the original multi-atlas labels using only 30 training examples. Furthermore, the manual corrections used as training examples were generated in a semi-automated method, which provides a time- and resource-efficient alternative to generate training data.

In the contrast-enhanced paired imaging dataset, we used mcSLANT to enforce similarity in automated segmentation between the pre- and post-contrast images. Again, we showed an improvement in performance of 0.05 DSC when the training data is a combination of manually traced examples and paired contrast-enhanced data. Despite the modest increase in overall DSC, this is an important issue in modern image processing. With increasing demand for larger sample size in deep learning, the field will likely turn to clinically acquired medical data for examples rather that acquiring more research scans. Segmentation algorithms will likely need to be robust to inhomogeneities present in clinically acquired data, such as intravenous contrast, movement, noise, or artifacts. Similarly, volumetric estimates, both in research and in clinical applications, will need to be consistent and reproducible within the same subject despite acquisition parameters or throughout a longitudinal timeline. Here, we showed that mcSLANT decreases the reproducibility error within the same subjects by an order of magnitude, which will result in more reliable and reproducible image processing on clinical data.

Overall, we showed that TL is a powerful tool to improve accuracy on a new dataset. However, we also learned that by performing TL exclusively on new subjects, there is a decrease in performance on the original adult cohort. We showed that using a mixed dataset during training offers a tradeoff between the two, with a larger benefit seen in the new dataset than it was detracted from the original one. Although the methods of integrating new data in an existing model is an active area of research, here we showed that using heterogenous training examples from manual and augmented labels can largely improve performance.

## 5. ACKNOWLEDGMENTS

This research was supported by NSF CAREER 1452485, NIH grants 1R01EB017230 (Landman), T32-EB021937 (NIH/NIBIB), and T32-GM007347 (NIGMS/NIH), and AHA grant 19PRE34380969. This research was conducted with the support from and the Charlotte and Donald Test Fund. This study was in part using the resources of the Advanced Computing Center for Research and Education (ACCRE) at Vanderbilt University, Nashville, TN. This project was supported in part by ViSE/VICTR VR3029 and the National Center for Research Resources, Grant UL1 RR024975-01, and is now at the National Center for Advancing Translational Sciences, Grant 2 UL1 TR000445-06. We gratefully acknowledge the support of NVIDIA Corporation with the donation of the Titan X Pascal GPU used for this research. The de-identified datasets used for the analysis described were obtained from the Research Derivative (RD), database of clinical and related data. The imaging dataset(s) used for the analysis described were obtained from ImageVU, a research repository of medical imaging data and image-related metadata. ImageVU and RD are supported by the VICTR CTSA award (ULTR000445 from NCATS/NIH) and Vanderbilt University Medical Center institutional funding. ImageVU pilot work was also funded by PCORI (contract CDRN-1306-04869).